\documentclass[a4paper,fleqn]{cas-sc}
\usepackage{soul}
\usepackage[sort&compress,numbers]{natbib}
\usepackage{lineno}
\usepackage{setspace}
\usepackage{multirow}

\usepackage{cuted}

\def\tsc#1{\csdef{#1}{\textsc{\lowercase{#1}}\xspace}}
\tsc{WGM}
\tsc{QE}
\tsc{EP}
\tsc{PMS}
\tsc{BEC}
\tsc{DE}

\begin{document}
\let\WriteBookmarks\relax
\def\floatpagepagefraction{1}
\def\textpagefraction{.001}
\shorttitle{Theoretical Prediction of Electron Mobility in Birhodanine Crystals and their Sulfur Analogues}
\shortauthors{Melo Neto \textit{et~al}.}

\title [mode = title]{Theoretical Prediction of Electron Mobility in Birhodanine Crystals and their Sulfur Analogues}

\author[1]{Carlos Alberto Moreira de Melo Neto}
\author[1]{Marcelo Lopes Pereira Junior}
\author[1,2]{Luiz Antonio Ribeiro Junior}
\author[1]{Dem\'etrio Ant\^onio da Silva Filho}
\cormark[1]
\ead{dasf@unb.br}

\cortext[cor1]{Corresponding Author}

\address[1]{Institute of Physics, University of Bras\'ilia, Bras\'ilia, Brazil.}
\address[2]{PPGCIMA, Campus Planaltina, University of Bras\'{i}lia, 73345-010, Bras\'{i}lia, Brazil.}

\begin{abstract}
    Molecular crystals compose the current state of the art when it comes to organic-based optoelectronic applications. Charge transport is a crucial aspect of their performance. The ability to predict accurate electron mobility is needed in designing novel organic semiconducting materials. In the present work, the Semi-Classical Marcus (SCM) and Marcus-Levich-Jortner (MLJ) hopping models are employed to numerically describe the charge mobility in six distinct birhodanine-like crystals. These materials were recently used in n-channel organic transistors as electron transporting layers. Results have revealed that the MLJ approach predicts electron mobilities in good agreement with the experiment, whereas SCM underestimates this parameter. Remarkably, we found for one of the birhodanine derivatives studied here average electron mobility of 0.14 cm$^2$ V$^{-1}$s$^{-1}$, which agrees with the one reported in experimental investigations. Moreover, it was identified that the MLJ approach presents a strong dependency on external reorganization energy. For SCM, a change in the reorganization energy value has a small impact on mobility, while for MLJ it impacts the average electron mobility that exponentially decays by increasing the external reorganization energy. Importantly, we highlight the primary source of the differences in predicting the electron mobility presented by both approaches, providing useful details that will help the selection of one of these two models for study different species of organic molecular crystals.
\end{abstract}




\maketitle
\doublespacing

\section*{Introduction}
Charge transport in organic semiconductors is a key process behind the operation in most of their applications, including OLEDs \cite{OLED2,OLED1}, OFETs \cite{OFET1,OFET2}, and OPVs \cite{OPV2,OPV1}. Several works in literature were conducted by confronting experimental measurements with theoretical models in order to estimate charge carrier mobility and model the charge transport phenomena \cite{huang2,intro4,intro3,intro5,theo-exp,pereirajr_jmm_2017a,pereirajr_jmm_2017b,pereirajr_pccp_2019,pereirajr_jpcc_2019a,pereirajr_jpcc_2019b,pereirajr_jmm_2019,pereirajr_sm_2019,pereirajr_scirep_2020,ribeirojr_PCCP_2017a,ribeirojr_PCCP_2017b,ribeirojr_2016,leal_jmm_2017}. Despite the progress achieved in this field in the past decade, there are still crucial challenges to overcome \cite{intro1,intro2}. From the experimental point of view, one of the main challenges is to obtain air-stable semiconductors for the electron transport, once electron mobility in organic-based materials usually decays rapidly after a period in exposure to air. In OFETs, for instance, there is a need for n-channel (electron-conducting) organic semiconductors with performance comparable to p-channel (hole conducting) materials, in order to promote their improvement \cite{newman_CM}.

Recently, birhodanines and their sulfur analogs --- rich electron acceptor molecular crystals --- were reported as high-performance n-channel transistors with impressive air stability in a study conducted by Iijima and coworkers \cite{base}. In their work, systematic experimental investigations were performed on a series of thin-film transistors (TFTs) based on various birhodanines to study their charge transport efficiency and molecular packing trend. The employed materials were 3,3'-dialkyl- 5,5'-bithiazolidinylidene-2,2'-dione-4,4'-dithiones (OS-R) and their sulfur analogues, 3,3'-dialkyl-5,5'- bithiazolidinylidene-2,4,2',4'-tetrathiones (SS-R), where R = Me, Et, Pr, and Bu. Their findings revealed that sulfur atoms impact molecular packing and the performance of the transistors. The SS-R crystals show characteristic tilted stacking structures attributed to the pronounced intermolecular S-S interactions. On the other hand, OS-R crystals have the ordinary herringbone structure owing to the reduced intermolecular interactions. Moreover, their results showed that SS-R TFTs exhibit better performance than the OS-R transistors as a consequence of the elongation of the alkyl chain length. Importantly, among all the birhodanine derivatives studies, SS-Pr exhibited remarkable stability even after air exposure for three months. Since birhodanine crystals have presented interesting traits in developing more efficient organic-based optoelectronic applications, the charge transport mechanism in these materials should be further understood to promote their broad usage.       

Herein, the Semi-Classical Marcus (SCM) and Marcus-Levich-Jortner (MLJ) hopping models are employed to theoretically describe the charge mobility in six distinct birhodanine-like crystals. Our numerical protocol is based on comparing these different theoretical methods for obtaining electronic properties in these materials. The results are contrasted with experimental data reported in reference \cite{base} to help the understanding of the procedures on simulating electron mobility for these recently developed high-performance electron-transport materials. Our findings showed that the MLJ approach predicts electron mobilities in good agreement with the experiment. For one of the birhodanine derivatives studied here, we found average electron mobility of 0.14 cm$^2$ V$^{-1}$s$^{-1}$, which agrees with the one reported in experimental investigations \cite{base}. Importantly, the SCM approach underestimates this parameter. 

\section*{Methods}
\begin{figure}[pos=ht]
	\centering
	\includegraphics[width=\linewidth]{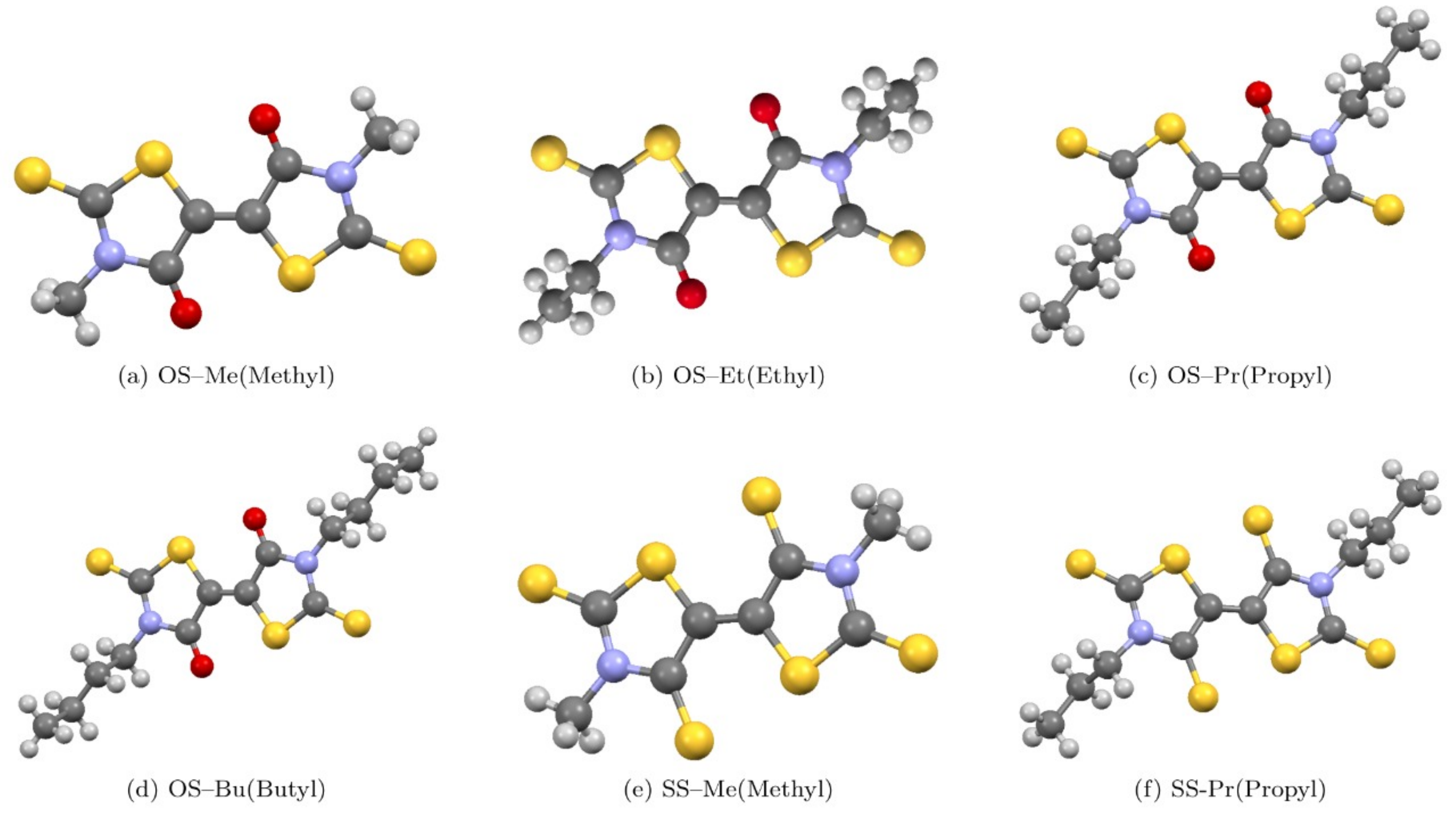}
	\caption{Diagrammatic representation of the birhodanine derivatives considered here: (a) OS--Methyl, (b) OS--Ethyl, (c) OS--Propyl, (d) OS--Butyl, (e) SS--Methyl, and (f) SS--Propyl.}
	\label{fig:mols}
\end{figure}

Motivated by the recent achievements on air-stable birhodanine-like crystals, we have performed a theoretical investigation of electron mobility in the systems presented in Figure \ref{fig:mols}. As demonstrated by Wetzelaer \textit{et al.} \cite{difusion}, in the thermal equilibrium regime, the electron mobility in disordered semiconductors can be estimated by the Einstein relation as follows: 
\begin{equation}
	\label{difusion}
	\displaystyle \frac{D}{\mu_e}=\frac{k_BT}{q}, 
\end{equation} 
where $D$ is the diffusion, $\mu_e$ the electron mobility, $k_B$ the Boltzmann's constant, $T$ the Temperature, and $q$ the elementary charge. As one can define the diffusion as $D=K_{et}r^2/2n$, with $K_{et}$ being the transfer rate of electrons from donors (D) to acceptors (A), $r$ the distance between two sites, and $n$ the dimension of the system ($n=1$ for our systems), equation \ref{difusion} can be rewritten as
\begin{equation}
	\displaystyle \mu_e = K_{et} \frac{ q r^2}{2k_B T}.
\end{equation}

To obtain an estimate for $\mu_e$ it is necessary to evaluate the transfer rates theoretically. The Semi-Classical Marcus equation \cite{marcus} has been the standard method to calculate $K_{et}$ and has the following form
\begin{equation}
	\label{eq.marcus}
	K_{et}^{Marcus}=\frac{\pi|H_{AD}|^2}{\hslash\sqrt{\pi\lambda k_BT}}\exp\left(\frac{-(\Delta G^0+\lambda)^2}{4\lambda k_BT}\right),
\end{equation}
where $\Delta G^0$ is the free energy difference between two sites, $\lambda$ is the reorganization energy ---  \textit{i.e.} the sum of internal ($\lambda_{in}$) and external ($\lambda_{ex}$) reorganization energies ---, and $H_{AD}$ is the electronic coupling term between LUMO levels of each molecule. Albeit widely used in calculating the charge transfer rates for organic materials, the Marcus theory usually underestimates these rates \cite{mlj-marcus}.

A more realistic way of obtaining the transfer rates is by Marcus-Levich-Jortner equation \cite{mlj,mlj2}, which includes quantum corrections for the Marcus equation taking into account the quantum nature of most active vibrational modes in molecular reorganization as follows 
\begin{equation}
	\label{eq.mlj}
	K_{et}^{MLJ} = \frac{\pi|H_{AD}|^2}{\hslash\sqrt{\pi\lambda_{ex} k_BT}} \sum_{\nu=0}^{\infty} e^{-S} \frac{S^{\nu}}{\nu!} \nonumber \exp\left(-\frac{(\Delta G^0+\lambda_{ex}+\nu\hslash\omega_{eff})^2}{4\lambda_{ex} k_BT}\right), 
\end{equation}
where $\nu$ is the quantum number of the \textit{i}th normal mode, $S=\lambda_{in}/\hslash\omega_{eff}$, and $\omega_{eff}$ is the effective frequency.

\begin{figure}[pos=ht]
	\centering
	\includegraphics[width=\linewidth]{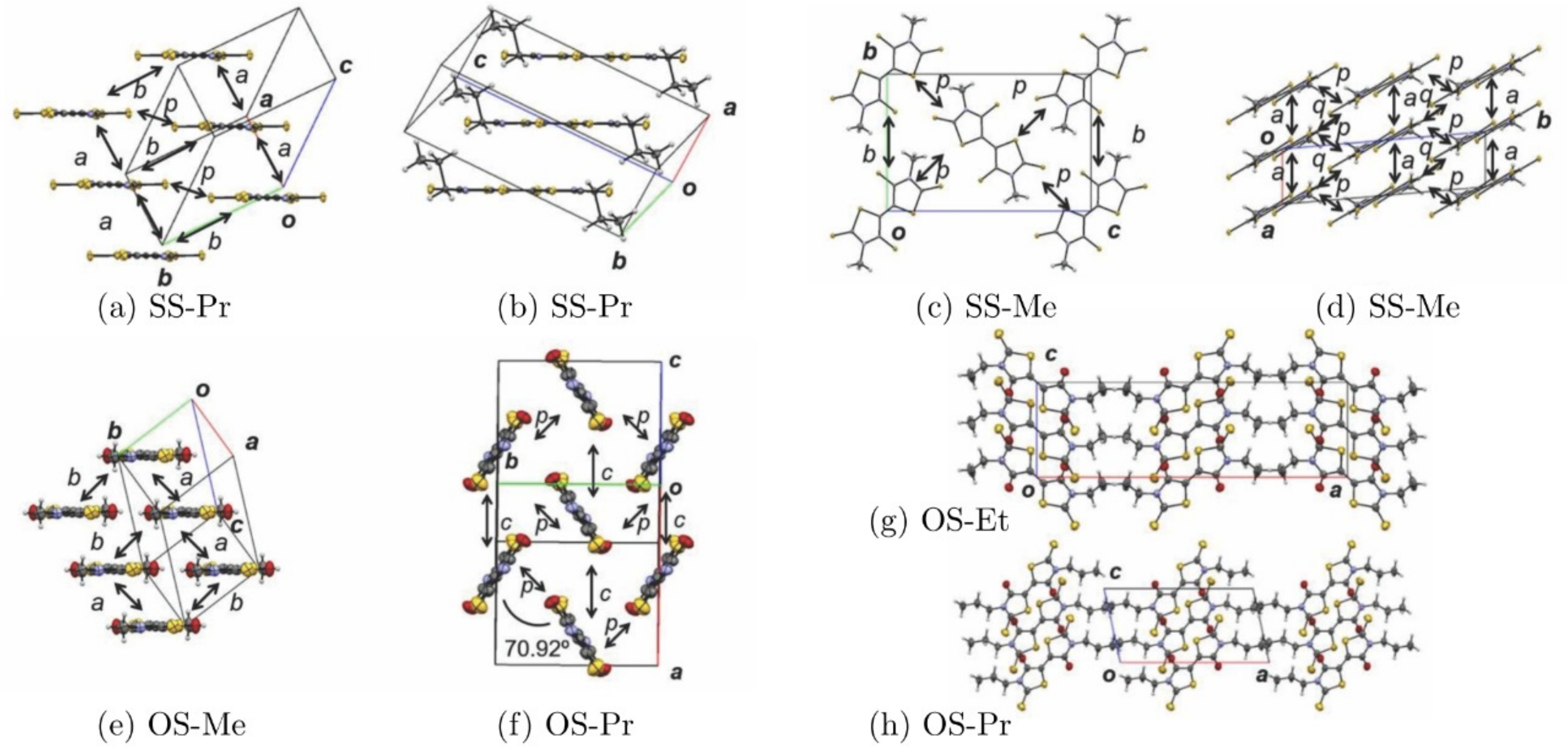}
	\caption{Diagrammatic representations of the possible hopping directions for studied crystals. For the sake of clarity, some cells are highlighted in two distinct orientations: (a-b) SS--Propyl, (c-d) SS--Methyl, (2) OS--Methyl, (f) OS--Propyl, (g) OS--Ethyl, and (h) OS--Propyl. This figure was adapted from reference \cite{base}.}
	\label{fig:directions}
\end{figure}

To apply the methods described above, optimized geometrical structures are necessary to perform vibrational analysis and electronic properties calculations of the chosen molecular systems. For this purpose, it was applied \textit{Gaussian 09} \cite{g09} software suite with DFT/CAM-B3LYP \cite{camb3} functional along with 6-31+g(d) \cite{6-31} as the basis set. Different possible dimmers were selected from the X-Ray structure \cite{base} to simulate the various possible pathways that charge hopping can happen. Only directions with coupling larger than 1 meV (which is roughly the accuracy of our estimation for this parameter) were considered in our analysis. In table \ref{tab:coup} it is shown the hopping for all directions. These directions are illustrated in Figures \ref{fig:directions}.

\begin{table}[pos=ht]
	\setlength{\tabcolsep}{3pt}
	\centering
	\caption{Electronic coupling (meV) absolute value for the different directions in each crystal.}
	\begin{tabular}{ccccccc}
		\hline\noalign{\smallskip}
		Direction &  OS-Me & OS-Et & OS-Pr & OS-Bu & SS-Me & SS-Pr \\
		a         &  26.03 & ----- & ----- & ----- & 103.27*&  4.51 \\
		b         &   9.01 & ----- & ----- & ----- &  1.22 & 40.14 \\
		c         &  ----- & 77.57 &  0.14 & 67.77 & ----- & ----- \\
		p         &  ----- & 34.55 & 19.37 & 17.99 & 79.82* & 20.14 \\
		q         &  ----- & ----- & ----- & ----- & 31.11* & ----- \\
		\noalign{\smallskip}\hline
	\end{tabular}
	\label{tab:coup}
\end{table}

With that in hand, python codes were implemented to find Huang-Rhys \cite{huang1,huang2,huang3} factor, effective frequencies, and reorganization energies from vibrational analysis to solve the equations for the transfer rates and electron mobility. The effective frequency can be found accounting the contribution from all frequencies from each normal mode and is associated to Huang-Rhys factor, according to the following expression \begin{equation}
	\displaystyle \omega_{eff} =\frac{\sum_{i=1}^{N} S_i\omega_i}{\sum_{i=1}^{N} S_i},   
\end{equation}
where $S_i$ is the Huang-Rhys factor which can be found by the vibrational analysis through all normal modes \textit{i} using the following equation
\begin{equation}
	\displaystyle S_i = \frac{\Delta Q_i^2 \mu_i \omega_i}{2\hslash}.
\end{equation}
In the equation above, $\Delta Q$ is the projection along with each normal mode of the geometrical displacement associated with the change in the charged state, \textit{i.e.}, from neutral to negatively charged state, $\mu_i$ is the reduced mass associated to the frequency $\omega_i$ of the \textit{i}th mode.

The total reorganization energy ($\lambda$) is a sum of internal ($\lambda_{in}$) and external ($\lambda_{ex}$) reorganization energies, where the internal reflects changes in geometry between neutral and charged states. This parameter was estimated in two ways: i) employing the four-point method \cite{4point}; and by adding up the contributions of each frequency mode to the changes in geometry, which is shown that both methods generate similar results \cite{reorg-vib}.

The four-point method is calculated using the energy of geometry optimizations from neutral $E_0^{0}$ and charged $E_{-}^{-}$ states, and single point energy calculations from charged state with neutral state geometry $E_{-}^{0}$, and neutral state with charged state geometry $E_{0}^{-}$ by the following expression:
\begin{equation}
	\displaystyle \lambda_{in}^{4p} = E_{0}^{-} - E_{0}^{0} + E_{-}^{0} - E_{-}^{-}.
\end{equation}
The reorganization energy can also be calculated accounting the contributions from the vibrational modes and the Huang-Rhys factors to the conformation of the molecules as follows
\begin{equation}
	\displaystyle \lambda_{in}^{vib} = \sum_{i=1}^N S_i\hslash\omega_i.
\end{equation}
For $\lambda_{ex}$ was used a value of 4 meV as a medium of various similar materials on the literature about organic semiconductors \cite{ext-reorg}.

The electronic couplings were calculated using the CATNIP (Charge Transfer Integral Package) \cite{catnip} software, that applies the transfer integral method \cite{coup1,coup2}, which calculates the transfer integral between orbital levels from two molecules. For electrons is considered LUMO--LUMO interactions, although in cases where the difference between LUMO and LUMO$+$1 are too close from each other LUMO$+$1 must be considered in the electron couplings. 

\section*{Results}
Our analysis begins by calculating effective frequency and reorganization energies from isolated molecules for each crystal. For the systems analyzed here, the average frequency is 780 cm$^{-1}$, the lowest is 765.346 cm$^{-1}$ for SS--Pr and the highest 796.672 cm$^{-1}$ for OS--Me. The equilibrium structures of single molecules were computed for ground and cation states in order to calculate reorganization energies from two methods, the four-point method and the contributions from normal modes vibrations, where results showed close enough to be considered equal. The OS--R systems accounted for higher reorganization energy than SS--R ones as it can be seen in Table \ref{tab:1}.
\begin{table}[pos=ht]
	\centering
	\caption{Effective frequencies in cm$^{-1}$, four point method ($\lambda_{in}^{4p}$) and vibrational contributions to ($\lambda_{in}^{Vib}$) reorganization energies in eV.}
	\label{tab:1}
	\begin{tabular}{lcccc}
		\hline\noalign{\smallskip}
		& $\omega_{eff}$ & $\lambda_{in}^{4p}$ & $\lambda_{in}^{Vib}$ \\
		\noalign{\smallskip}\hline\noalign{\smallskip}
		OS--Methyl & 796.672 & 0.632 & 0.645  \\
		OS--Ethyl  & 782.109 & 0.636 & 0.640  \\
		OS--Propyl & 777.088 & 0.636 & 0.641  \\
		OS--Butyl  & 775.288 & 0.636 & 0.643  \\
		SS--Methyl & 783.213 & 0.472 & 0.471  \\
		SS--Propyl & 765.346 & 0.480 & 0.479  \\
		\noalign{\smallskip}\hline
	\end{tabular}
\end{table}

\begin{figure}[pos=ht]
	\centering
	\includegraphics[width=\linewidth]{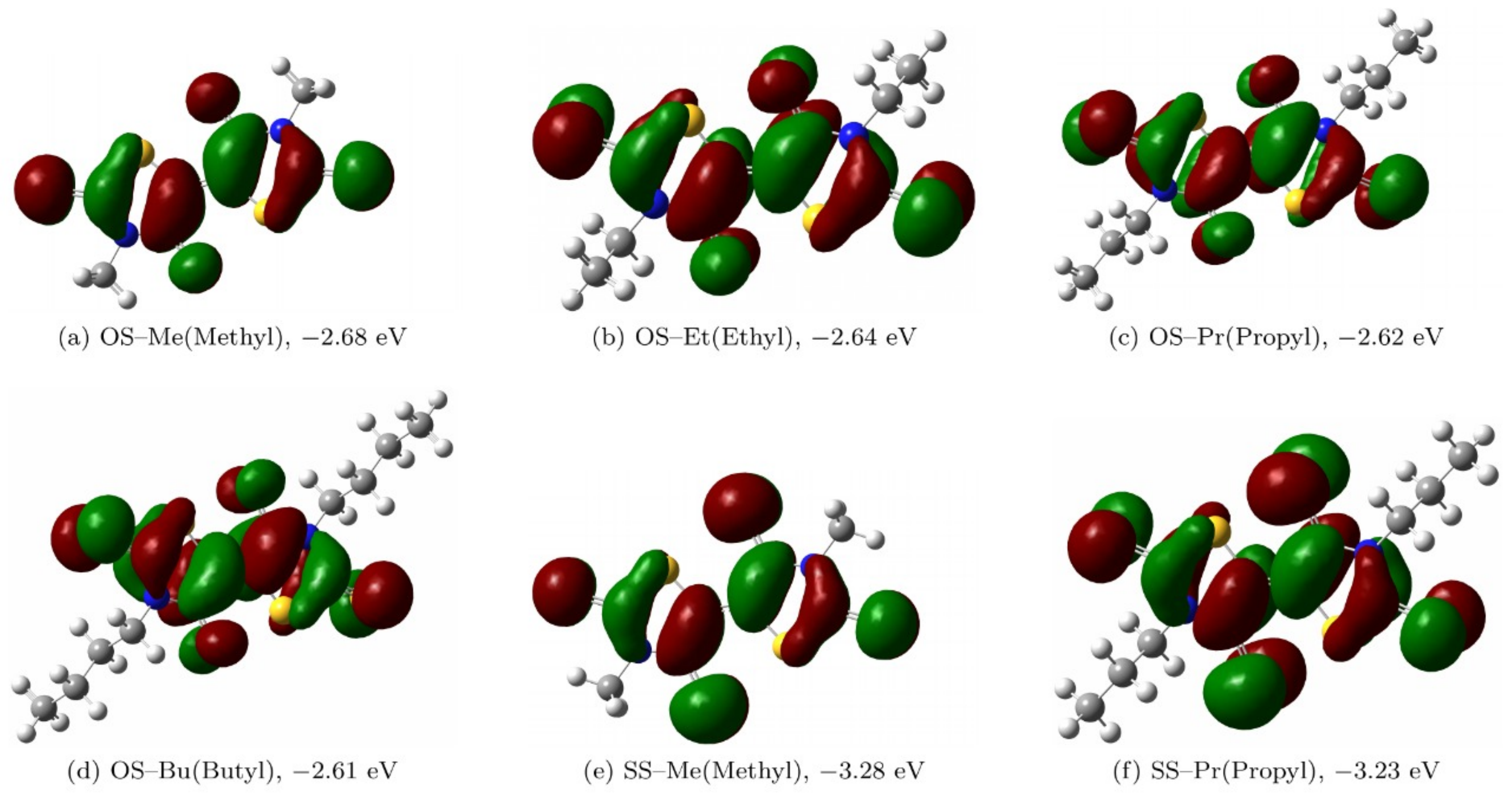}
	\caption{LUMO levels configuration for studied molecules: (a) OS--Me(Methyl), (b) OS--Et(Ethyl), (c) OS--Pr(Propyl), (d) OS--Bu(Butyl), (e) SS--Me(Methyl), and (f) SS--Pr(Propyl).}
	\label{fig:lumo}
\end{figure}

With the electronic couplings presented in Table \ref{tab:2} and the parameters previously acquired, SCM and MLJ rates can be computed. Since our objective is to investigate the transfer of electrons, the electronic coupling must be between LUMO levels from each dimer of the molecules considered here (see Figure \ref{fig:lumo}). Importantly, the LUMO level is uniformly distributed over the molecules and it present similar shape among them. The results for orbital energies and optical gaps are in Table \ref{tab:energy}. Comparing electron mobilities computed with SCM and MLJ equations it was found that the second is in better agreement with the experimental data. For SS--Pr crystal with MLJ our result for $\mu_e$ was $0.25 cm^2V^{-1}s^{-1}$ and the experimental measure is $0.24 cm^2V^{-1}s^{-1}$ \cite{base} while with SCM rate the charge mobility estimated was $0.03 cm^2V^{-1}s^{-1}$, one order of magnitude lower, underestimating $\mu_e$ as expected.
\begin{table}[pos=ht]
	\setlength{\tabcolsep}{4pt}
	\centering
	\caption{Energy levels HOMO (H), HOMO-1(H-1), LUMO(L), LUMO+1(L+1) and optical gaps in eV.}
	\label{tab:energy}
	\begin{tabular}{lccccc}
		\hline\noalign{\smallskip}
		& $E_{L}$ & $E_{L+1}$ & $E_{H}$ & $E_{H-1}$ & Gap  \\
		\noalign{\smallskip}\hline\noalign{\smallskip}
		OS--Methyl & -2.68 & -0.85 & -8.02 & -8.41 & 5.33  \\
		OS--Ethyl  & -2.64 & -0.82 & -7.95 & -8.36 & 5.32  \\
		OS--Propyl & -2.62 & -0.80 & -7.93 & -8.34 & 5.32  \\
		OS--Butyl  & -2.61 & -0.79 & -7.92 & -8.33 & 5.32  \\
		SS--Methyl & -3.28 & -1.13 & -7.95 & -8.18 & 4.67  \\
		SS--Propyl & -3.23 & -1.07 & -7.88 & -8.11 & 4.65  \\
		\noalign{\smallskip}\hline
	\end{tabular}
\end{table}

\begin{figure}[pos=ht]
	\centering
	\includegraphics[width=0.6\linewidth]{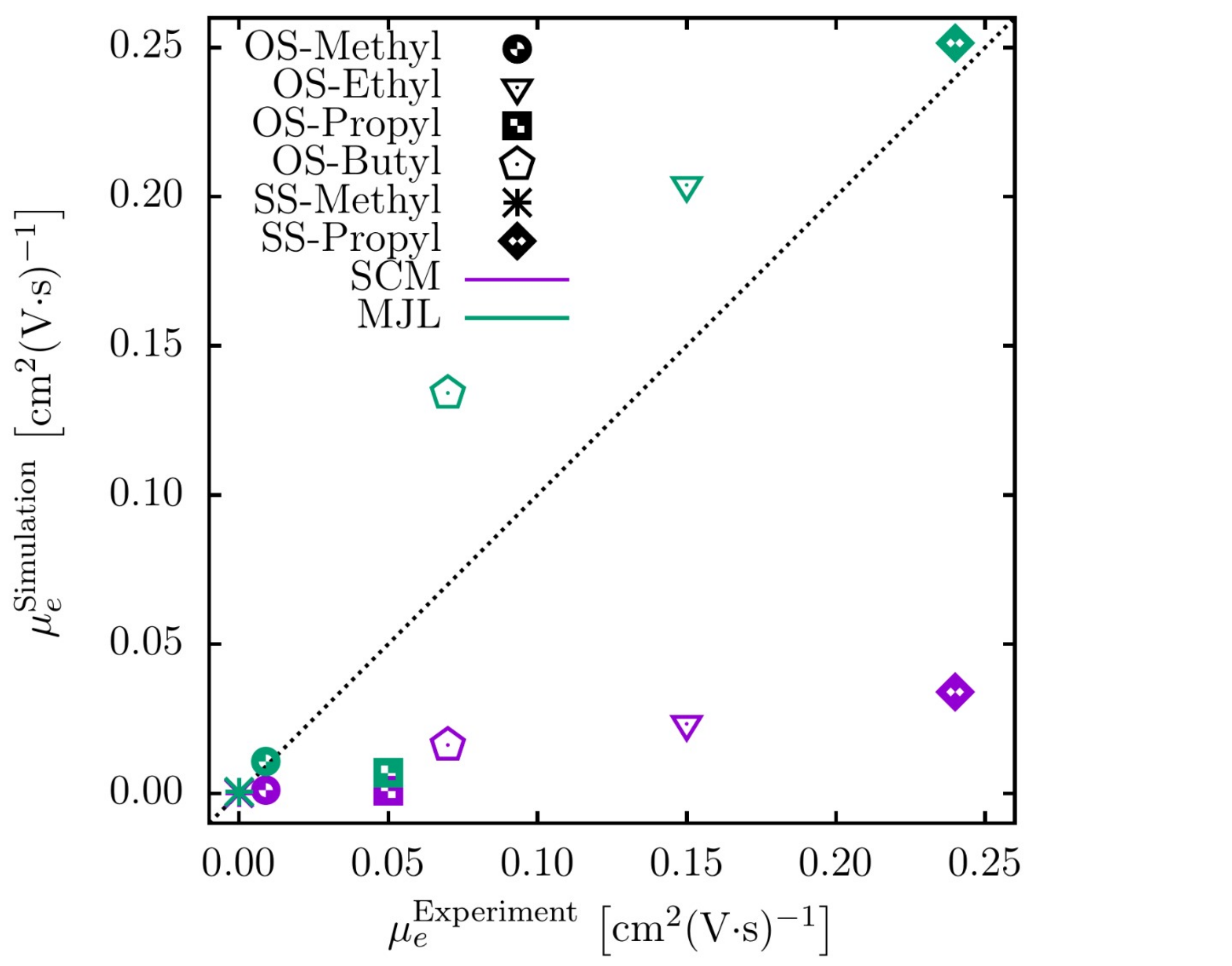} 
	\caption{\normalsize Mobilities calculated with SCM and MLJ rates versus experimental mobility \cite{base}.}
	\label{fig:mobil}
\end{figure}

Figure \ref{fig:mobil} contrasts the electron mobilities calculated here using SCM and MLJ with the experimental values reported in reference \cite{base}. As a guide for the eye, the dashed line marks the ideal relationship between theory and experiment. In this figure, one can note that just in the OS--Methyl and SS--Methyl cases there is a good concordance with the experiment for both approaches. In regimes of considerably small electron mobilities, SCM and MLJ are equivalent in predicting the electron mobilities in organic molecular crystals. As a general trend, MLJ tends to present mobility values close to the experimental ones, while SCM tends to underestimate them. For higher electron mobilities, the SS-Propyl molecule has presented the best agreement with the experimental data ($0.25$ cm$^2$V$^{-1}$s$^{-1}$ for MLJ and $0.24$ for the experimental mobility \cite{base}).         

\begin{figure}[pos=ht]
	\centering
	\includegraphics[width=0.6\linewidth]{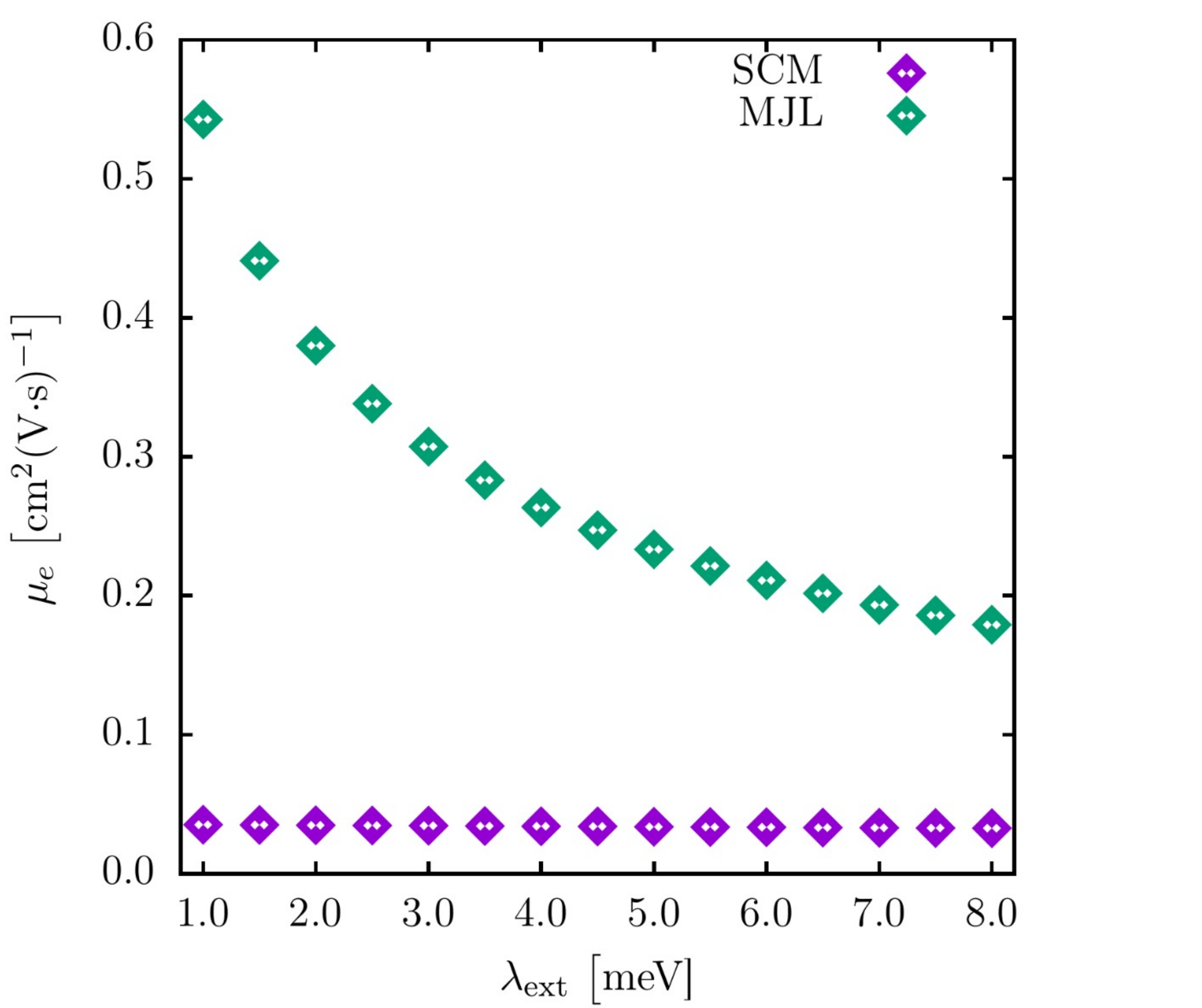} 
	\caption{Mobilities as function of the external reorganization energy ($\lambda_{Ext})$ for the SS--Pr as the representative case, once it is the crystal with higher electron mobility.}
	\label{fig:mob2}
\end{figure}

Here, it was identified that the MLJ approach presents a strong dependency on the external reorganization energy, and to the best of our knowledge, there is no reliable method to estimate this property. While for SCM approach the reorganization energy is the sum of external and internal energies, for MLJ the internal reorganization energy appears as a dependency on the effective frequency and on the external relaxation energy directly in the equation. This correlation with the external reorganization energy can be seen in Fig. \ref{fig:mob2}, where for pure SCM, a change in reorganization energy value has small impact on mobility, while for MLJ it impacts the average electron mobility that exponentially decays by increasing the external reorganization energy. In this figure, the interval was set based on known external reorganization energies for organic materials and the equations evaluated for SS--Pr as a representative case, once it is the crystal with higher electron mobility. Despite this dependence on the value of reorganization energy, our results shows that MLJ approach better estimates the mobility values when contrasted to SCM one, considering the external reorganization energy as a medium from other similar organic materials.

Based on our findings, the best material was SS--Pr due to its high electron mobility when compared to the other ones studied here as was also reported in the experimental data \cite{base}. Comparing the mobilities with electronic coupling, (see Table \ref{tab:2}) highest couplings not always leads to best mobilities as in OS--Et and OS--Bu, the best material (SS--Pr) was not the one with the highest coupling. Thus one can conclude that estimate only the electronic coupling is not sufficient in finding materials with better average mobilities.
\begin{table}[pos=ht]
	\setlength{\tabcolsep}{4pt}
	\centering
	\caption{Electronic couplings ($H_{AD}$) in meV, transfer rates ($K_{et}$) in $s^{-1}$, electron mobilities ($\mu_e$) in $cm^2V^{-1}s^{-1}$ and distance between monomer centers (R) in \AA}
	\label{tab:2}
	\begin{tabular}{lclccccc}
		& H$_{AB} $& $K_{et}^{Marcus}$ & $K_{et}^{MLJ}$ & $\mu_e^{Marcus}$ & $\mu_e^{MLJ}$ & $\mu_e^{Exp}$ & R\\
		\hline\noalign{\smallskip}
		OS--Me & $26.03$ & $2.45\times10^{10}$  & $2.43\times10^{11}$  & $0.0011$ & $0.0107$  & $0.009$ & $4.73$ \\
		OS--Et & $77.57$  & $2.29\times10^{11}$  & $2.01\times10^{12}$  & $0.0233$ & $0.2039$  & $0.15$  & $7.20$ \\
		OS--Pr & $19.37$ & $1.41\times10^{10}$  & $1.18\times10^{11}$  & $0.0008$ & $0.0069$  & $0.05$  & $5.44$ \\
		OS--Bu & $67.77$ & $1.70\times10^{11}$  & $1.40\times10^{12}$  & $0.0162$ & $0.1342$  & $0.07$  & $6.99$ \\
		SS--Me & $1.22$  & $3.41\times10^{8}$   & $2.82\times10^{13}$  & $6.7\times10^{-5}$  & $5.5\times10^{-4}$ & $7.3\times10^{-5}$ & $9.99$ \\
		SS--Pr & $40.14$ & $3.41\times10^{11}$  & $2.52\times10^{12}$  & $0.0340$ & $0.2515$  & $0.24$  & $7.13$ \\
		\noalign{\smallskip}\hline
	\end{tabular}
\end{table}

\section*{Conclusion}
In summary, the Semi-Classical Marcus and Marcus-Levich-Jortner hopping models were employed to theoretically describe the charge mobility in distinct birhodanine-like crystals. Here, it was showed how the MLJ equation appears to be more accurate than the SCM equation for electron transfer rates in high-performance materials. Results have revealed that the MLJ approach predicts electron mobilities in good agreement with the experiment, whereas SCM underestimates this parameter. Despite the dependency on external reorganization energy, our results shown MLJ calculated values closer to the experimental data ($0.25$ cm$^2$V$^{-1}$s$^{-1}$ for MLJ and $0.24$ for the experimental mobility \cite{base}). Mobilities calculated with the pure SCM equation are underestimated for most cases. Importantly, a change in the reorganization energy value has a small impact on mobility when the SCM approach was employed, while for MLJ it impacts the average electron mobility that exponentially decays by increasing the external reorganization energy. The material with the highest estimated mobility was SS--Pr as in experimental data \cite{base}, although for electronic coupling was not the highest one. For OS--Et and OS--Bu was computed higher couplings. Therefore, one can conclude that estimate only couplings should not be the best approach in finding better electron-transport materials.  

\printcredits

\bibliographystyle{cas-model2-names}

\bibliography{cas-refs}

\end{document}